\documentclass[conference]{IEEEtran}
\IEEEoverridecommandlockouts
\usepackage{cite}
\usepackage{amsmath,amssymb,amsfonts}
\usepackage{algorithmic}
\usepackage{graphicx}
\usepackage{textcomp}
\usepackage{xcolor}
\usepackage{graphicx}
\usepackage{multirow}
\usepackage{booktabs} 
\usepackage{adjustbox}
\usepackage{etoolbox}
\usepackage[a4paper, total={184mm,239mm}]{geometry}
\usepackage{hyperref}
\hypersetup{
    hypertex=true,
    colorlinks=true,
    linkcolor=black,     
    anchorcolor=black,   
    citecolor=black,    
    urlcolor=blue      
}

\def\BibTeX{{\rm B\kern-.05em{\sc i\kern-.025em b}\kern-.08em
    T\kern-.1667em\lower.7ex\hbox{E}\kern-.125emX}}

\begin{document}

\title{CIM-Based Parallel Fully FFNN Surface Code High-Level Decoder for Quantum Error Correction%
\thanks{A conference version of this work has been accepted
for presentation at the 2025 Design, Automation \& Test
in Europe Conference (DATE).
DOI: \href{https://doi.org/10.23919/DATE64628.2025.10993142}
{10.23919/DATE64628.2025.10993142}.}%
}
\author{
    \IEEEauthorblockN{
        Hao Wang\textsuperscript{1*}, 
        Erjia Xiao\textsuperscript{1*},
        Wenbo Mu\textsuperscript{2}
        Songhuan He\textsuperscript{2}, 
        Zhongyi Ni\textsuperscript{1}, 
        Lingfeng Zhang\textsuperscript{1}, 
        Xiaokun Zhan\textsuperscript{3}, \\
        Yifei Cui\textsuperscript{2},
        Jinguo Liu\textsuperscript{1}, 
        Cheng Wang\textsuperscript{2+}, 
        Zhongrui Wang\textsuperscript{4+}, 
        Renjing Xu\textsuperscript{1+}
    }
    \IEEEauthorblockA{
        \centering
        \parbox{\textwidth}{\centering
            \textsuperscript{1}Hong Kong University of Science and Technology (Guangzhou), \\ 
            \textsuperscript{2}University of Electronic Science and Technology of China, \\
            \textsuperscript{3}Harbin Institute of Technology, \\ 
            \textsuperscript{4}Southern University of Science and Technology
        }
    }
    \IEEEauthorblockA{\centering
        wangch87@uestc.edu.cn; wangzr@sustech.edu.cn; renjingxu@hkust-gz.edu.cn
    }
}




\maketitle

\begin{abstract}
Due to the high sensitivity of qubits to environmental noise, which leads to decoherence and information loss, active quantum error correction(QEC) is essential. Surface codes represent one of the most promising fault-tolerant QEC schemes, but they require decoders that are accurate, fast, and scalable to large-scale quantum platforms. In all types of decoders, fully neural network-based high-level decoders offer decoding thresholds that surpass baseline decoder-Minimum Weight Perfect Matching (MWPM), and exhibit strong scalability, making them one of the ideal solutions for addressing surface code challenges. However, current fully neural network-based high-level decoders can only operate serially and do not meet the current latency requirements (below 440 ns). To address these challenges, we first propose a parallel fully feedforward neural network (FFNN) high-level surface code decoder, and comprehensively measure its decoding performance on a computing-in-memory (CIM) hardware simulation platform. With the currently available hardware specifications, our work achieves a decoding threshold of 14.22\%, and achieves high pseudo-thresholds of 10.4\%, 11.3\%, 12\%, and 11.6\% with decoding latencies of 197.03 ns, 234.87 ns, 243.73 ns, and 251.65 ns for distances of 3, 5, 7 and 9, respectively. The impact of hardware parameters and non-idealities on these results is discussed, and the hardware simulation results are extrapolated to a 4K quantum cryogenic environment.
\end{abstract}


\section{Introduction}
Quantum computing holds significant promise as a solution to complex problems that classical computers struggle to solve \cite{harrow2017quantum}. Unfortunately, qubits are extremely sensitive to their environment, making them prone to decoherence, which can lead to the loss of stored information\cite{varsamopoulos2019comparing}. To counter this, quantum error correction (QEC) is essential.

QEC involves two critical steps: encoding and decoding. Numerous robust encoding techniques have been developed \cite{bolt2016foliated, terhal2015quantum, ryan2021realization}, among which the surface code\cite{fowler2012surface} stands out due to it is easy to implement on physical platforms. However, the surface code decoding process is constrained by stringent decoder-threshold and decoder-delay requirements, meaning the decoder must be both highly accurate and fast enough to keep pace with the QEC cycle (e.g., the microsecond($\mu$s) time scale)\cite{battistel2023real}. Moreover, future quantum systems will consist of large-scale arrays of qubits \cite{fowler2012surface}\cite{gidney2021factor}, necessitating that decoders meet three key design constraints: accuracy, latency, and scalability \cite{das2022afs}.

\begin{figure}[ht!]
\centering
\includegraphics[width=1.0\linewidth]{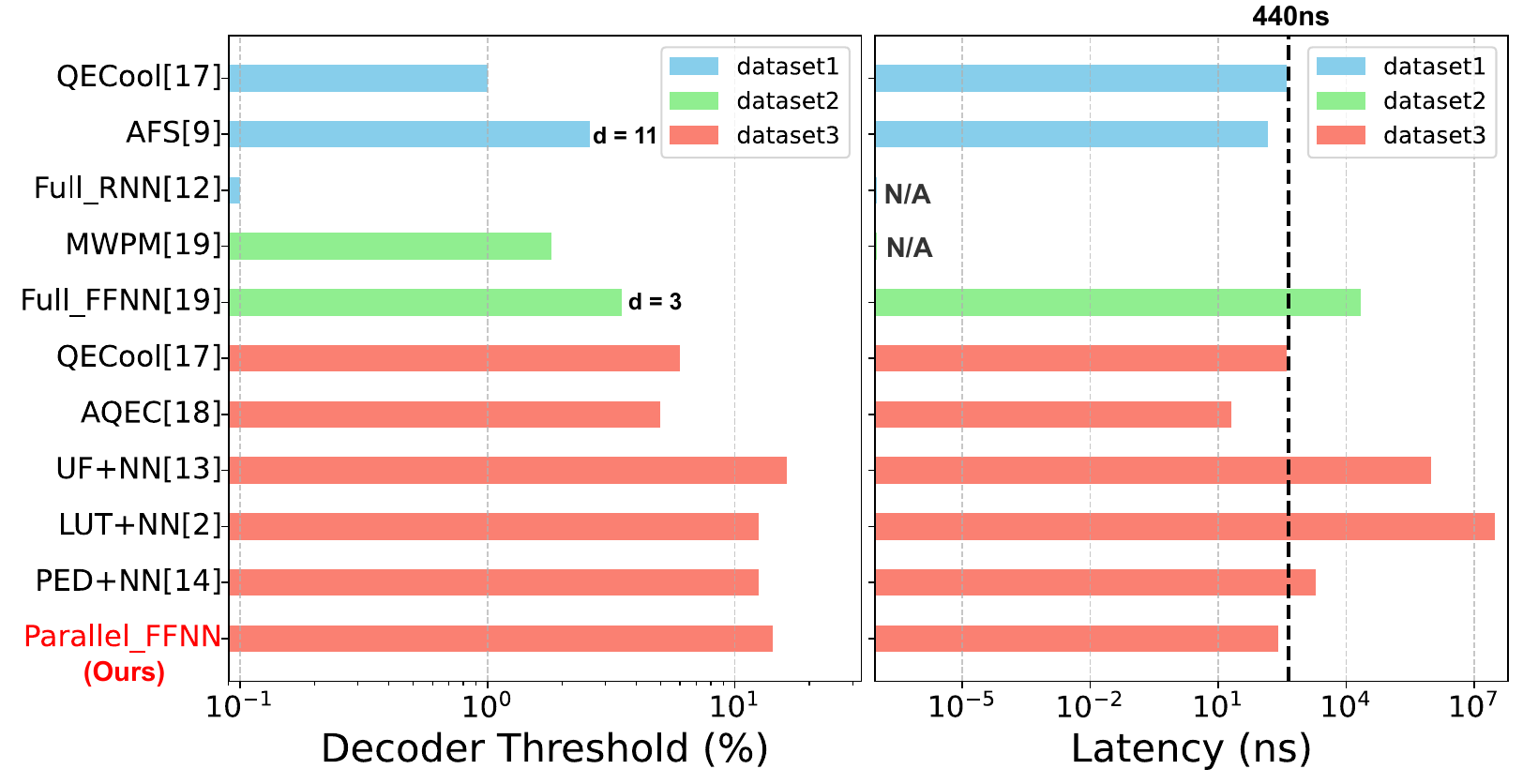}
\caption{\small {Comparison of distance = 9 (except for special annotations) surface code decoder performances. The datasets—Dataset1, Dataset2, and Dataset3—represent the data from the Circuit-Level Noise Model, the Modified Depolarizing Noise Model, and the Depolarizing Noise Model, respectively.}}
\vspace{-10pt}
\label{fig.1}
\end{figure}

Several decoders, such as Union Find (UF) \cite{das2022afs}, Look-Up Table (LUT) \cite{das2022lilliput}, Neural Network (NN)-based \cite{baireuther2018machine,marcotte2023cryogenic,meinerz2022scalable,overwater2022neural,cao2023qecgpt}, Minimum-Weight Perfect Matching (MWPM) \cite{google2023suppressing}, and modified MWPM (mod-MWPM) \cite{ueno2021qecool}\cite{holmes2020nisq+}, have been employed for surface code decoding. Although LUT\cite{das2022afs}, UF\cite{das2022lilliput}, and mod-MWPM-based \cite{ueno2021qecool}\cite{holmes2020nisq+} decoders have demonstrated decoding latencies below 440ns (the most stringent quantum decoding latency \cite{overwater2022neural}) on small code distances (distance $\leq 9$) in 4k qubit simulations, they reduce decoding accuracy (threshold), falling short of the most commonly used baseline decoder (MWPM), as shown in Fig.~\ref{fig.1}.

Neural network(NN)-based decoders have garnered considerable attention due to their potential to surpass MWPM in terms of decoding performance and scalability. However, they face certain challenges. NN-based decoders can be categorized into two types: low-level decoders (containing a single neural network module) and high-level decoders (comprising a simple decoder and a classifier module) \cite{varsamopoulos2019comparing}. Classifier module typically use neural networks. The simple decoder module can either incorporate neural networks\cite{baireuther2018machine}\cite{marcotte2023cryogenic}\cite{bhoumik2021efficient}  or function with non-NN rules\cite{varsamopoulos2019comparing}\cite{meinerz2022scalable}\cite{overwater2022neural}. High level decoders constituted by non-neural network-based simple decoder modules demonstrate superior decoding performance and allow for parallel execution with the classifier component, although they exhibit poor scalability. Conversely, fully neural network-based high level decoders exhibit enhanced scalability. However, existing implementations operate serially and have yet to achieve satisfactory outcomes in terms of decoding thresholds and latency. Thus, the key challenge for NN-based decoders is to develop scalable, parallel fully neural network-based high-level decoders that maintain performance and latency on decoding tasks.

To address this challenge, we propose a novel fully-parallel high-level decoder comprising a simple decoder and classifier module, both built using a two-layer Feedforward Neural Network(FFNN). Our decoder achieves high decoding thresholds(14.22\%) on small-distance tasks under depolarizing noise models. Moreover, using the MNSIM 2.0 \cite{zhu2023mnsim} simulator, we first demonstrate quantum surface code decoding on code distances 3 to 9 using a computing-in-memory (CIM) hardware architecture, achieving below-440ns latency on available hardware specifications at 300K ambient temperature. Finally, we discuss hardware non-idealities and extend the results to 4k quantum cryogenic environments. Fig.~\ref{fig.1} compares our work with existing decoders(distance = 9). Our contributions are summarized as follows:

\begin{itemize}
    \item \textbf{Fully FFNN High-Level Decoder:} We propose a parallel fully NN-based high-level decoder, constructed with a two-layer FFNN, achieving high decoding thresholds(14.22\%) on code distances 3 to 9 under depolarizing noise models.
    
    \item \textbf{CIM-Based Decoder:} We first demonstrate a CIM-based surface code fully NN-based high-level decoder using the MNSIM 2.0 simulator, analyzing the impact of various hardware specifications and temperature conditions on decoder latency.
    
    \item \textbf{Low Hardware Decoder Latency:} We achieve below-440ns latency on code distances 3 to 9 with the proposed decoder, using available hardware specifications and CIM architecture.
\end{itemize}

\section{Related Works}
\subsection{NN-Based low level decoder}
Neural network-based decoders, which generally outperform MWPM(baseline) in terms of decoding performance, are typically categorized into low-level and high-level decoders. Low-level decoders are generally more scalable. For example, \cite{gicev2023scalable} utilizes a CNN to achieve a decoding threshold of 0.138 for the distances ranging from 9 to 513 under the depolarizing noise model, which is currently the largest-scale decoder demonstrated.

\subsection{NN-Based High level decoder}
High-level decoders typically offer superior decoding performance with lower implementation complexity, but scalability and decoding latency remain challenges. For instance, \cite{meinerz2022scalable} employs Union Find (UF) for the simple decoder module of High-level decoder, achieving a decoding threshold of 0.162 for code distances from 3 to 127, with decoding latencies in the millisecond(\text{ms}) range. Similarly, \cite{varsamopoulos2019comparing} uses a look-up table (LUT) as the simple decoder, achieving a pseudo-threshold of 12.4\% on distance 9 with a hardware latency of 31.34 ms on a CPU. Additionally, \cite{overwater2022neural} employs the Pure Error Decoder (PED) as the simple decoder, reaching a pseudo-threshold of 12.49\% on distance 9. The neural network component's decoding latency was measured as 87.6 ns on FPGA and 14.3 ns on ASIC, though the PED latency was not reported.

\begin{figure}[ht!]
\centering
\includegraphics[width=1.0\linewidth]{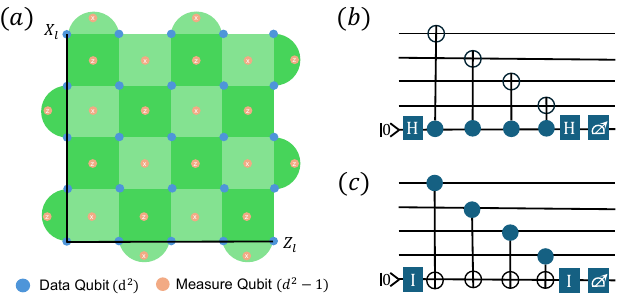}
\caption{\small {Distance = 5 Surface Code Structure. (a) Two-Dimensional Schematic of the Surface Code; (b) X-Ancillas Circuits; (c) Z-Ancillas Circuits.}}
\label{fig.2}
\vspace{-10pt}
\end{figure}

Fully NN-based high-level decoders offer improved scalability but have yet to meet the desired decoding performance and latency. \cite{baireuther2018machine} introduced an LSTM-based simple decoder but did not report decoding latency or threshold. \cite{bhoumik2021efficient} implemented FFNN/CNN-based simple decoders, achieving thresholds of 3.5\% and 3.4\% for distances 3 and 7, surpassing MWPM baseline (1.81\%). But these results were under a modified depolarizing noise model, and even the simplest FFNN (distance 3) required 21 µs on a CPU. \cite{marcotte2023cryogenic} demonstrated a CIM-based RNN decoder, achieving a 0.1\% pseudo-threshold on a circuit-level noise model for distance 3, though decoding latency was not reported.

\begin{figure*}[ht!]
\centering
\includegraphics[width= 0.95\linewidth]{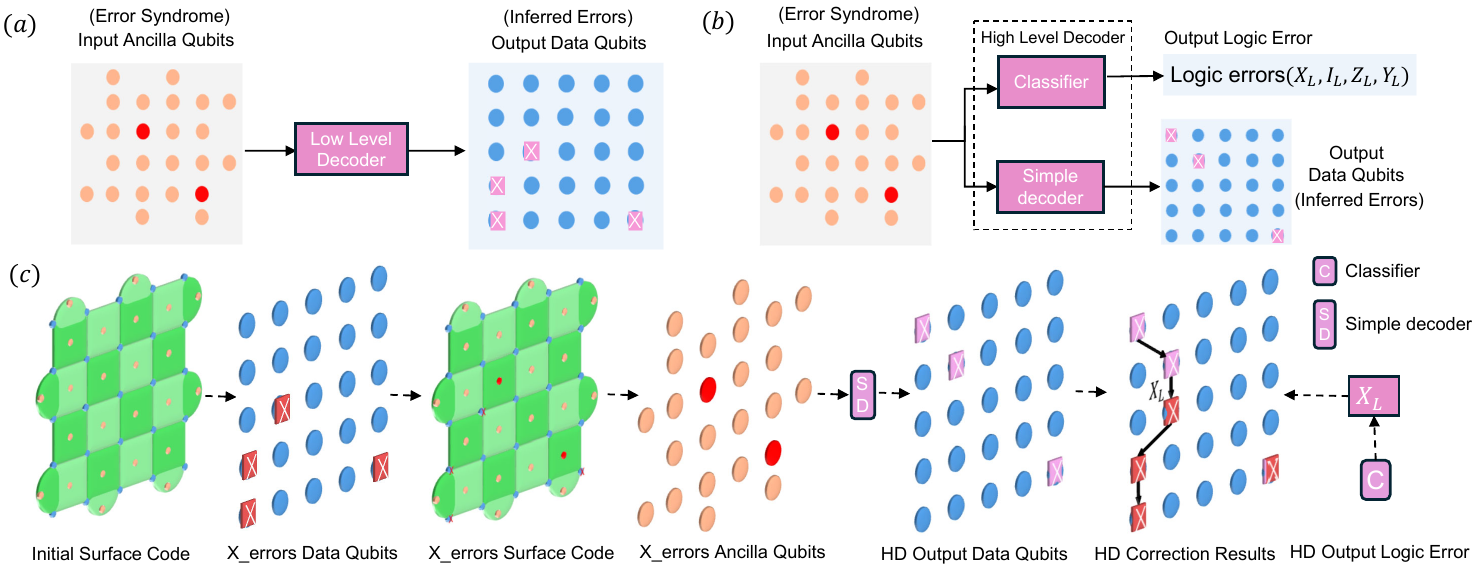}
\caption{\small {Schematic of NN-Based Neural Network Decoders. (a) Low-Level Decoder; (b) High-Level Decoder (HD); (c) Example of a High-Level Decoder for Distance = 5 surface code.}}
\label{fig.3}
\vspace{-10pt}
\end{figure*}

\section{Decoder Design}
\subsection{Suface code and decoder}
A brief description of the surface code is provided here; for more detailed information, please refer to \cite{fowler2012surface}\cite{overwater2022neural}. As shown in Fig.~\ref{fig.2} (a), the surface code is a fault-tolerant encoding constructed from $d \times d$ data qubits and $d^2 - 1$ measurement qubits (ancilla qubits). The data qubits redundantly represent a single logical qubit, where $d$ denotes the code distance. The logical qubit states are defined by a pair of anti-commuting logical observables, $X_l$, $Z_l$, and $Y_l = X_l Z_l$.  Since directly measuring the data qubits would cause quantum collapse and result in the loss of stored information, ancilla qubits are employed to build quantum circuits for information readout. The quantum circuits constructed from ancilla qubits and their neighboring data qubits come in two types: X-ancillas and Z-ancillas, which are used to correct Z errors and X errors, respectively, as illustrated in Fig.~\ref{fig.2} (b) and (c). The \textbf{error syndrome} is the combination of quantum information obtained from ancilla qubits. 

Surface code decoding involves using the error syndrome to predict the state of the data qubits (if an error has occurred) and subsequently implementing error correction. However, since multiple combinations of data qubit errors can produce the same error syndrome, the decoder can only output the most likely correction operation, making this computation NP-hard \cite{ueno2021qecool}. MWPM is the most commonly used quantum decoder\cite{google2023suppressing}\cite{google2021exponential}\cite{krinner2022realizing}.

\subsection{NN-Based Surface code decoder}
Neural network-based decoders can be divided into low-level and high-level decoders. This section provides a detailed explanation. As shown in Fig.~\ref{fig.3} (a), the low-level decoder uses the error syndrome (ancilla qubits) as input and the data qubits(inferred errors) as output. Generally, the low-level decoder only consists of a neural network module. For a low-level decoder to be successful, it must accurately correct the errors on each data qubit. Low-level decoder works well with datasets that have very low physical error rates but performs poorly at higher physical error rates. This is because when the number of occurring errors exceeds \( \frac{d}{2} \), different combinations of data qubit errors may produce the same error syndrome. As illustrated in Fig.~\ref{fig.3} (a) and (b), The different combinations of data qubit errors decoded result in the same error syndrome.

The high-level decoder also takes the error syndrome as input and comprises two modules: a simple decoder and a classifier. The classifier is typically implemented using a neural network, while the simple decoder can be either neural network-based \cite{baireuther2018machine}\cite{marcotte2023cryogenic}\cite{bhoumik2021efficient} or non-neural network-based \cite{varsamopoulos2019comparing}\cite{meinerz2022scalable}\cite{overwater2022neural}. Fully neural network-based architectures offer better scalability, but these decoders must first obtain the most likely result from the NN-based simple decoder and then use the NN-based classifier for logical error correction. This process is executed serially, introducing additional latency. Non-neural network-based simple decoders map each error syndrome to a specific combination of data qubit errors, enabling them to run in parallel with the classifier. However, scalability is limited by the non-NN-based simple decoder, as the delay grows exponentially with increasing distance \cite{das2022afs}.

Fig.~\ref{fig.3} (c) illustrates the current approach using a high-level decoder with a non-NN-based simple decoder. When the real X-errors in the data qubits cause errors in the ancilla qubits (input error syndrome), a lookup table (LUT) method from \cite{varsamopoulos2019comparing} is applied as the simple decoder, producing HD output data qubits. These HD output data qubits, along with the real X-errors in the data qubits, result in the logical error \(X_L\). During the decoding process, the classifier simultaneously determines the most probable logical error (in this case, \(X_L\)). After obtaining the HD output data qubits, the logical error is corrected. Here, \(X_L\), \(Y_L\), \(Z_L\), and \(I_L\) represent  the logical qubit errors in \(X_l\), \(Y_l\), \(Z_l\) and no any logical error, respectively.

\begin{figure*}[ht!]
\centering
\includegraphics[width=0.95\linewidth]{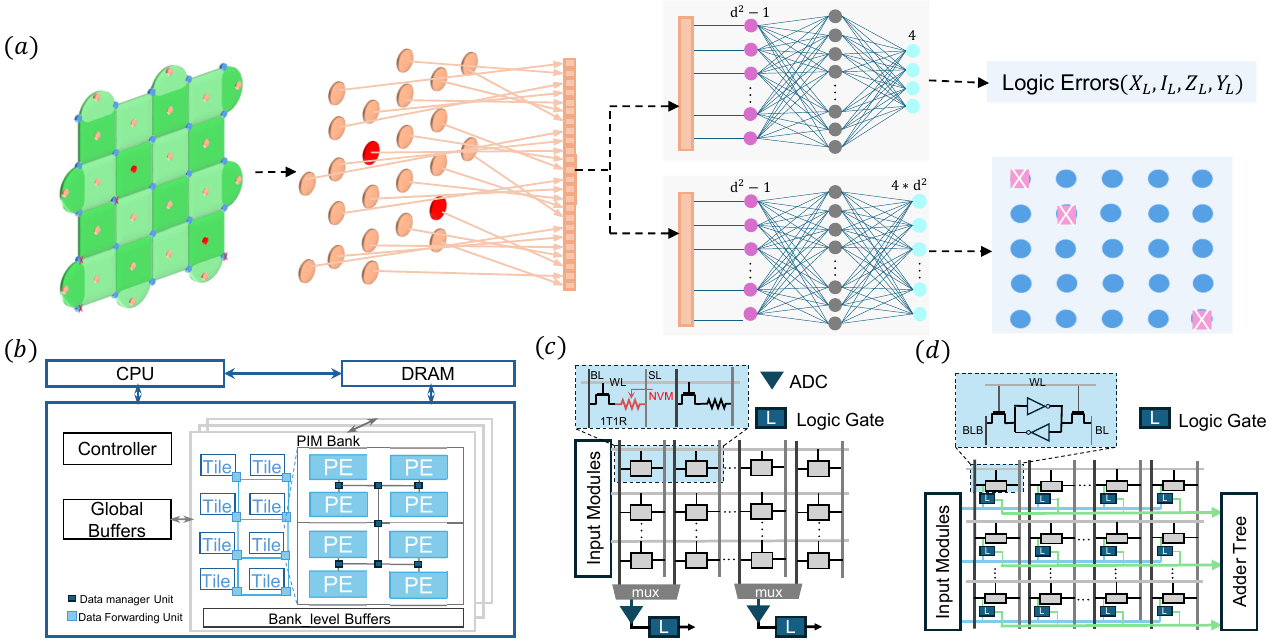}
\caption{\small {Structure of the Parallel fully NN-Based High-Level Decoder and CIM Simulator.(a) Our Parallel fully NN-Based High-Level Decoder; (b) MNSIM 2.0 Simulator Architecture(Adapted from\cite{zhu2023mnsim}); (c) NVM-Based Array for MNSIM; (d) SRAM-Based Array for MNSIM.}}
\label{fig.4}
\vspace{-10pt}
\end{figure*}

\subsection{Parallel fully NN-Based high-level decoder}
Inspired by \cite{overwater2022neural} and \cite{bhoumik2021efficient}, we first constructe a fully FFNN-based high-level decoder with parallel execution, as illustrated in Fig.~\ref{fig.4} (a). To achieve parallel execution, we employed the PED-based simple decoder from \cite{overwater2022neural} to generate training data for our NN-Based simple decoder, transforming the problem into a fixed-category classification task where each input error syndrome corresponds to a unique error combination on the data qubits.

This approach offers three major advantages:  
\textit{(i)} \textbf{Improved scalability}. The PED component is solely used for generating training data, no longer contributing to the decoder's execution latency. The only components impacted by distance expansion are the two-layer FFNN currently in use.  
\textit{(ii)} \textbf{Reduced execution latency}. In our CPU-based evaluation, the PED component incurs a decoding latency of $\sim 2 \mu$s for distance-9 tasks, increasing exponentially with distance. In contrast, the neural network can leverage advanced acceleration platforms to achieve decoding latencies below 440ns.  
\textit{(iii)} \textbf{Maintained high decoding performance}. Since the actual decoding rules match the PED-based high-level decoder, our decoder maintains the superior decoding performance of the PED-based high-level decoder, outperforming the 10\%.

\textbf{FFNN Neural Network.} The Feedforward Neural Network (FFNN), also known as the fully connected network (FCN), is employed in our decoder in two distinct forms: one for the simple decoder and one for the classifier, both illustrated in Fig.~\ref{fig.4} (a). For the classifier, the network input is the error syndrome (of size $d^2-1$), and the output corresponds to the four possible logical errors. Adjusting the size of the hidden layer further improves prediction performance. For the simple decoder, the network input is also the error syndrome (of size $d^2-1$), but the output represents the error combination on the data qubits. Since each data qubit can be in one of four states (X, Y, Z error, or no error), the output layer requires $4 \times d^2$ neurons. By tuning the hidden layer size, we can nearly achieve a 100\% match with the PED's output. Both networks utilize ReLU as the activation function.

\begin{figure*}[ht!]
\centering
\includegraphics[width=1.0\linewidth]{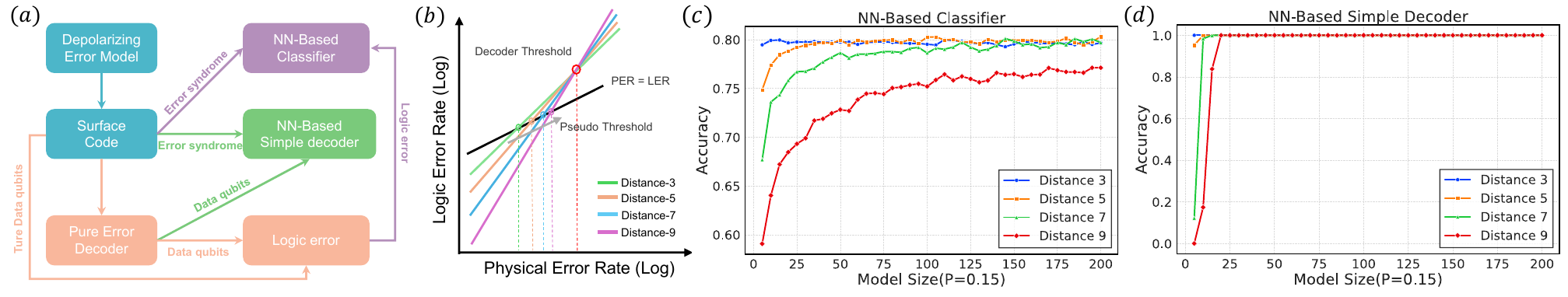}
\caption{\small {(a) Decoder Training Process; (b) Decoding Thresholds and Pseudo-Thresholds; (c) Impact of NN Model Size on the Classifier; (d) Impact of NN Model Size on the Simple Decoder.}}
\label{fig.5}
\vspace{-10pt}
\end{figure*}

\subsection{CIM-based high-level decoder}
We present a fully FFNN-based high level decoder with code distances ranging from 3 to 9 using the comprehensive CIM simulation platform MNSIM 2.0 \cite{zhu2023mnsim}. MNSIM\cite{zhu2023mnsim}\cite{xia2017mnsim}, as one of the most comprehensive computing in-memory(CIM) simulators to date, has been applied in numerous related studies \cite{xie2024dypim}\cite{song2024fully}\cite{lyu2024memristive}. We employ the latest version 2.0 \cite{zhu2023mnsim}. Fig.~\ref{fig.4} (b) illustrates the hardware architecture of the MNSIM 2.0 simulator, while Fig.~\ref{fig.4} (c) and (d) depict CIM-based architectures leveraging NVM (Non-volatile Memristor) and SRAM (Static Random Access Memory), respectively. The simulator enables complete simulation from training to hardware testing for both CNN and FFNN algorithms under the CIM framework, including modules for pooling layers, ReLU functions, convolution layers, and FFNN layers. All of our experiments were conducted using the MNSIM 2.0 architecture.

\section{Evaluation}

\subsection{Experimental preparation}
\textbf{Dataset.} In our experiment, we use the depolarizing noise model to generate the dataset, as detailed in \cite{varsamopoulos2019comparing} and \cite{overwater2022neural}. The procedure for generating the dataset is described in \cite{varsamopoulos2019comparing}. According to the proposed guidelines, the minimum training set sizes for surface code decoding with distances of 3, 5, 7, and 9 should be 256, \(2 \times 10^5\), \(3 \times 10^6\), and \(2 \times 10^7\), respectively. Open-source code is provided in \cite{overwater2022neural}, and following the prescribed rules, we generated training sets for both the NN-Based simple decoder and NN-Based classifier with a fixed physical error rate of 0.15. The test set evaluates the decoder's performance across a range of physical error rates from 0.03 to 0.3, and for each physical error rate, we generate test sets of the same size as the training sets for distances 3 to 9.

\textbf{Training.} Fig.~\ref{fig.5} (a) illustrates the process of generating our dataset and training the models. For the NN-Based simple decoder, the error syndromes from the surface code serve as the input to the network, while the predicted error combinations on the data qubits, as determined by the PED module, form the output. Similarly, for the NN-Based classifier, the error syndromes from the surface code serve as the network's input, but the output is the logical error, obtained by comparing the predicted data qubits error combinations from the PED module with the actual data qubits error combinations. Both neural network models are trained using the modules provided by MNSIM 2.0, with a training epoch count of 30, using the Adam optimizer and a learning rate of 0.001.

\textbf{Decoding Threshold.} One of the key metrics used to evaluate the decoder performance is the decoding threshold, which represents the fault tolerance of the decoder. As shown in Fig.~\ref{fig.5} (b), the colored lines depict the decoding performance of surface code decoders at different distances, while the black line represents the performance without surface code encoding (i.e., where the logical error rate equals the physical error rate). The physical error rate at which the decoder achieves a distance-independent logical error rate (indicated by the red circle) is referred to as the decoding threshold (Dth). A larger decoding threshold indicates a higher fault tolerance of the decoder. The intersections of the decoder performance curves at various distances with the black line correspond to the pseudo-thresholds (Pth), where the logical error rate begins to become lower than the physical error rate (indicated by other colored circles). The pseudo-thresholds are different for each surface code distance and serve as a measure of fault tolerance for decoders at the same distance.

\textbf{Decoding Latency.} Another crucial metric for assessing decoder performance is decoding latency, which generally needs to be less than 440ns \cite{overwater2022neural} to be applicable for future quantum computers. For serially executed high level decoders \cite{baireuther2018machine}\cite{marcotte2023cryogenic}\cite{bhoumik2021efficient}, decoding latency is the total time spent by both the simple decoder and the classifier. For parallel executed high level decoders (as in our work and \cite{varsamopoulos2019comparing}\cite{meinerz2022scalable}\cite{overwater2022neural}), decoding latency is determined by the maximum time taken between the simple decoder and the classifier.
\subsection{Neural network size}
In Section 3, we have already detailed the input layer and output layer sizes for the NN-Based simple decoder and NN-Based classifier. This section presents simulations that determine the size of the hidden layers. The size of the hidden layers is defined by the following equation:

\[
H(\text{size}) = n \times \text{distance}
\]

where \(n\) is an integer between 5 and 200. We perform simulations using the MNSIM 2.0 simulator, iterating \(n\) at intervals of 5. The results are shown in Fig.~\ref{fig.5}. Fig.~\ref{fig.5} (c) illustrates the relationship between training accuracy and \(n\) for the NN-Based classifier, while Fig.~\ref{fig.5} (d) presents the corresponding results for the NN-Based simple decoder. Due to the trade-off between network size and latency, compounded by the impact of hardware non-idealities, we opted for larger network sizes to enhance robustness. Although larger networks introduce higher latency, they provide better performance under non-ideal conditions. For example, in the NN-Based classifier with a network for distance 9, when \(n = 20\), the prediction accuracy in ideal conditions is 100\%, but it drops to 21\% under the influence of non-idealities. However, when \(n = 35\), the accuracy remains as high as 99.9\% even with non-idealities. Based on our experiments, the optimal \(n\) values for the NN-Based classifier are 20, 40, 60, and 80 for distances 3, 5, 7, and 9, respectively. For the NN-Based simple decoder, we selected \(n\) values of 5, 15, 25, and 35, respectively.

\begin{figure*}[ht!]
\centering
\includegraphics[width=1.0\linewidth]{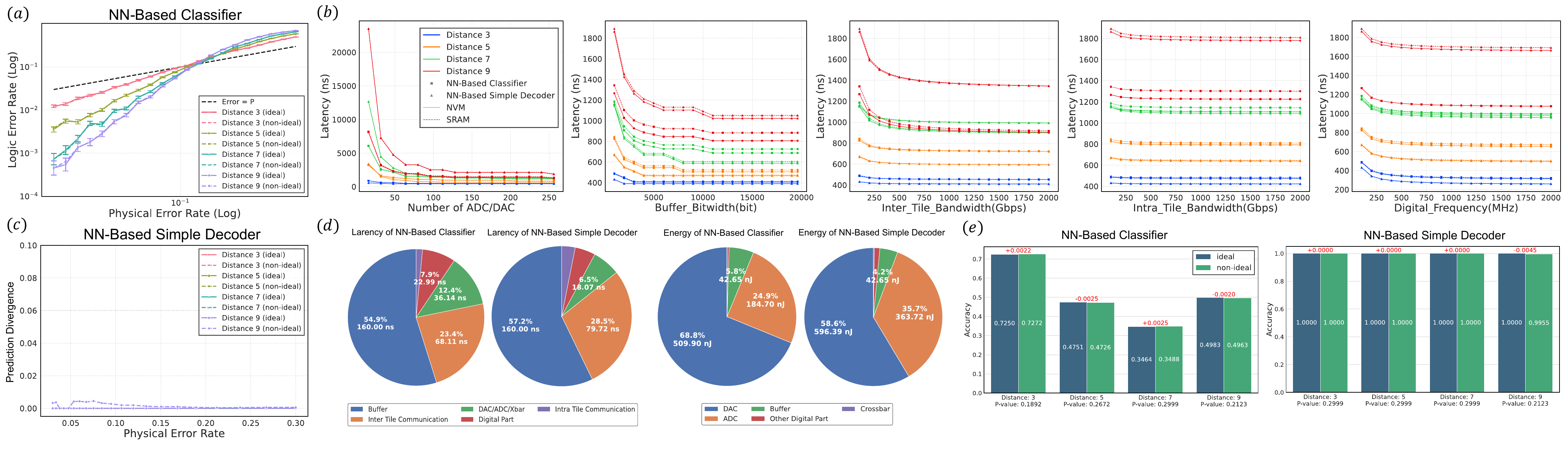}
\caption{\small {Decoder hardware performance. (a) NN-Based classifier results; (b) Relationship between 5 hardware parameters and latency; (c) NN-Based simple decoder results; (d) Proportion of each hardware parameter in total latency and energy; (e) Maximum impact of hardware non-idealities on results.}}
\label{fig.6}
\vspace{-10pt}
\end{figure*}

\subsection{experimental results}
\textbf{Decoding Threshold Results.} Following the previously mentioned training rules, we completed the training of two networks on MNSIM 2.0 and began testing the performance of the decoders. Fig.~\ref{fig.6} (a) shows the pure algorithm simulation (solid line) and the simulation results including hardware non-idealities (dashed line), with a 99\% confidence interval for the NN-based classifier. The figure indicates the decoding threshold and pseudo-decoding threshold at various distances. Our decoder achieved a high decoding threshold of 14.22\%, surpassing the 10\%. Additionally, for distances 3, 5, 7, and 9, we achieved pseudo-decoding thresholds of 10.4\%, 11.3\%, 12\%, and 11.6\%, respectively. Fig.~\ref{fig.6} (c) shows the simulation results of the NN-based simple decoder, which achieved nearly 100\% prediction similarity for PED results, demonstrating the feasibility of our method.

\textbf{Decoding Latency Results.} For decoding latency, we first needed to set the hardware parameters in MNSIM 2.0 that affect the runtime of the CIM architecture. In the experiment, we used data from ADCs and DACs proposed in \cite{liu202210gs} and \cite{caragiulo20212}. Additionally, we discussed the five most impactful hardware parameters, including Digital Frequency, Buffer Bitwidth, Inter-Tile Bandwidth, Intra-Tile Bandwidth, and Number of ADC/DACs. The results of these discussions are shown in Fig.~\ref{fig.6} (b) and Table~\ref{table.1}. Other hardware parameters used the default values in MNSIM 2.0. Digital Frequency represents the operating frequency of the digital circuits, which we set to a maximum of 2GHz, a common value. Inter-Tile Bandwidth and Intra-Tile Bandwidth represent the data bandwidth between tiles and within tiles (between PEs), respectively, as shown in Fig.~\ref{fig.4} (b). We set the maximum values to 256GBps (2048Gbps), based on data from the recently presented high-performance Dojo chip \cite{talpes2022dojo}. Buffer Bitwidth represents the bitwidth of data read/write from the buffer, and we set the maximum value to 19600 bits, following the Nvidia High Bandwidth Memory (HBM) chip \cite{nvidia_grace_hopper}. The Number of ADC/DAC represents the number of ADCs and DACs connected to a \(256\times256\)-sized crossbar, with a maximum value of 256. We also presented discussions on SRAM-Based CIM and NVM-Based CIM. We found that the NVM-Based CIM platform (where each NVM stores only 0/1 bit) outperformed the SRAM-Based CIM platform in terms of latency, and ultimately chose NVM-Based CIM for result presentation.

\begin{table}[htbp]
\centering
\caption{simulation hardware parameters}
\label{table.1}
\resizebox{\linewidth}{!}{
\begin{tabular}{l|cccc|cccc}
\toprule
\multirow{2}{*}{\textbf{Hardware parameters}} & \multicolumn{4}{c|}{\textbf{NN-Based Classifier}} & \multicolumn{4}{c}{\textbf{NN-Based Simple Decoder}} \\ \cmidrule(lr){2-5} \cmidrule(lr){6-9}
                                 & \textbf{d=3} & \textbf{d=5} & \textbf{d=7} & \textbf{d=9} & \textbf{d=3} & \textbf{d=5} & \textbf{d=7} & \textbf{d=9} \\ \midrule
\textbf{Digital Frequency (MHz)}  & 1500         & 1500         & 1500         & 1500         & 1500         & 1500         & 1500         & 1500         \\ 
\textbf{Buffer Bitwidth (bit)}    & 3000         & 8000         & 11000        & 11000        & 2000         & 4000         & 8000         & 11000        \\ 
\textbf{Inter Tile Bandwidth (Gbps)} & 1000      & 1000         & 1500         & 1500         & 1000         & 1000         & 1000         & 1500         \\ 
\textbf{Intra Tile Bandwidth (Gbps)} & 1000      & 1000         & 1000         & 1000         & 600          & 800          & 1000         & 1000         \\ 
\textbf{Number of ADC/DAC}        & 96           & 144          & 144          & 144          & 64           & 112          & 208          & 256          \\ 
\bottomrule
\end{tabular}
}
\end{table}

\vspace{-10px}

\begin{table}[htbp]
\centering
\caption{decoder hardware performance(nvm-based)}
\label{table.2}
\resizebox{\linewidth}{!}{
\begin{tabular}{l|c|cccc|cccc}
\toprule
\multirow{2}{*}{\textbf{Distance}} & \multirow{2}{*}{\textbf{Pth}} & \multicolumn{4}{c|}{\textbf{NN-Based Classifier}} & \multicolumn{4}{c}{\textbf{NN-Based Simple Decoder}} \\ \cmidrule(lr){3-6} \cmidrule(lr){7-10}
                          &                          & \textbf{Latency(ns)} & \textbf{Area(mm\(^2\))} & \textbf{Power(W)} & \textbf{Energy(nJ)} & \textbf{Latency(ns)} & \textbf{Area(mm\(^2\))} & \textbf{Power(W)} & \textbf{Energy(nJ)} \\ \midrule
\textbf{3}                & 10.40\%                  & 197.03  & 72.92   & 0.16  & 31.18   & 196.11  & 55.55   & 0.11  & 20.92   \\ 
\textbf{5}                & 11.30\%                  & 234.87  & 98.97   & 0.75  & 177.06  & 203.93  & 81.61   & 0.22  & 45.79   \\ 
\textbf{7}                & 12\%                     & 243.73  & 197.95  & 1.89  & 461.63  & 215.37  & 133.72  & 0.54  & 116.74  \\ 
\textbf{9}                & 11.60\%                  & 251.65  & 296.93  & 2.94  & 740.82  & 239.65  & 479.34  & 4.23  & 1018.12 \\ 
\bottomrule
\end{tabular}
}
\end{table}


\begin{table}[htbp]
\centering
\caption{comparison results(distance = 9)}
\label{table.3}
\resizebox{\linewidth}{!}{
\begin{tabular}{lccccccc}
\toprule
\textbf{Methods} & \textbf{Dth} &\textbf{Latency} & \textbf{Area} & \textbf{Power} & \textbf{Platform} & \textbf{Noise Model} & \textbf{Temperature}\\
                         
                          \midrule
\textbf{MWPM\cite{bhoumik2021efficient}}                & 1.81\%                  & -  & -   & -  & CPU  & Self-modified  & 300k      \\
\textbf{Fully\_FFNN\cite{bhoumik2021efficient}}                & 3.50\%                  & 21.6\(\mu s\)(d=3)  & -   & -  & CPU  & Self-modified  & 300k      \\
\midrule
\textbf{MWPM\cite{ueno2021qecool}}                & 2.90\%                  & -  & -   & -  & CPU  & Circuit-level  & 300k      \\
\textbf{QECool\cite{ueno2021qecool}}                & 1\%                  & 400ns  & 183.45mm\(^2\)   & 400.32\(\mu W\)  & SFQ  &  Circuit-level & 4k      \\
\textbf{AFS\cite{das2022afs}}                & 2.60\%                  & 150ns(d=11)  & -   & -  & -  & Circuit-level  & 300k      \\
\textbf{Fully\_RNN\cite{marcotte2023cryogenic}}                & 0.10\%                  & -  & -   & -  & CIM  & Circuit-level  & 300k      \\
\midrule
\textbf{QECool\cite{ueno2021qecool}}                & 6\%                  & 400ns  & 183.45mm\(^2\)   & 400.32\(\mu W\) &SFQ  & Depolarizing  & 4k      \\
\textbf{AQEC\cite{holmes2020nisq+}}                & 5\%                  & 19.8ns  & 329.46mm\(^2\)   & 3.88mW  & SFQ  & Depolarizing  & 4k      \\
\textbf{UF+NN\cite{meinerz2022scalable}}                & 16.20\%                  & \textgreater 1ms  & -   & -  & FPGA  & Depolarizing  & 300k      \\
\textbf{LUT+NN\cite{varsamopoulos2019comparing}}                & \textgreater 12.45\%                  & 31.34ms  & -   & -  & CPU  & Depolarizing  & 300k      \\
\textbf{PED+NN\cite{overwater2022neural}}                & \textgreater 12.49\%                  & 1.936\(\mu s\)  & -   & -  & CPU  & Depolarizing  & 300k      \\
\textbf{PED+NN\cite{overwater2022neural}}                & \textgreater 12.49\%                  & 5.312\(\mu s\)  & -   & -  & FPGA  & Depolarizing  & 300k      \\
\midrule
\textbf{Ours(Parallel\_FFNN)}                & \textbf{14.22\%}                  & \textbf{251.65ns}  & \textbf{479.34mm\(^2\)}   & \textbf{6.99W}  & \textbf{CIM}  & Depolarizing  & 300k      \\
\textbf{Ours(Parallel\_FFNN)}                & \textbf{14.22\%}                  & \textbf{221.07ns}  & \textbf{479.34mm\(^2\)}   & \textbf{3.98W}  & \textbf{CIM}  & Depolarizing  & 4k/0.4V\\
\bottomrule
\end{tabular}
}
\end{table}

\textbf{Hardware Performance of NVM-Based CIM Platform.} Table~\ref{table.2} shows the hardware performance achieved on the NVM-Based CIM platform under the hardware parameters listed in Table~\ref{table.1}, including latency, area, power, and energy consumption at various distances. For distances 3 to 9, our decoder showed latency below 440ns. For distance 9, we compared our results with other works, as shown in Table~\ref{table.3} and Fig.~\ref{fig.1}. Since \cite{overwater2022neural} did not report latency for the PED portion, we supplemented the latency results for FPGA (250MHz) and CPU using open-source code. Fig.~\ref{fig.6} (d) presents the impact of different hardware components in the CIM architecture on latency and energy consumption, providing further optimization directions. It specifically highlights the significant influence of Cryogenic ADC/DAC and Cryogenic HBM platforms on the NN-based quantum surface code decoder. This is also the first complete presentation of a surface code advanced decoder on a CIM architecture, along with a discussion of various influential hardware metrics.

\textbf{Hardware Non-Idealities.} MNSIM 2.0 takes into account three major non-ideal factors of memory devices: stuck-at-faults (SAFs), limited on/off ratio, and resistance variations. We used the default non-idealities provided by MNSIM 2.0, which have been shown to match the performance of real chips in \cite{yan20221} and \cite{liu202033}. Fig.~\ref{fig.6} (a) and (c) have already shown the impact of non-idealities during the entire decoding process, and Fig.~\ref{fig.6} (e) presents the maximum impact of hardware non-idealities on different distance tasks in the NN-based simple decoder and NN-based classifier. In the network sizes we proposed, the maximum impact of non-idealities was less than 0.5\%.


\textbf{4K Cryogenic Environment.} The above results were all based on temperature (300K). Although some works \cite{homulle2019cryogenic}\cite{van2020cryo} have discussed the hardware performance of mature transistor circuits from room temperature to cryogenic conditions, complete cryogenic CIM chips have not yet been demonstrated. The most relevant cryogenic CIM work is \cite{wang2021cryogenic}, which showed that latency decreased by $\sim 12.15\%$ as the temperature dropped from 300K to 4K, while energy consumption remained almost unchanged. However, by reducing the operating voltage from 0.9V to 0.4V, energy consumption decreased by $\sim~43\%$. We extended our work to the 4K cryogenic environment and 0.4V operating voltage using parameters provided in \cite{wang2021cryogenic}, with the results presented in Table~\ref{table.3}. 

\section{CONCLUSION}
We first present the parallel execution of a fully FFNN high-level surface code decoder and provide a comprehensive demonstration of a quantum surface code decoder performance under a computing-in-memory (CIM) architecture, exploring the impact of various hardware parameters on the decoder. With the currently achievable hardware specifications, our decoder attains a high decoding threshold of 14.22\%. Additionally, it achieves a decoding latency of less than 440 ns for decoding tasks ranging from distance 3 to 9. Future research could explore the decoder's performance at greater distances, attempt to optimize the execution architecture of CIM for quantum error correction (QEC), and develop training methods that require lower memory usage.

\begingroup
\footnotesize
\bibliographystyle{IEEEtran}
\bibliography{template-A4}
\endgroup

\end{document}